\documentclass[letterpaper, 11pt]{article}
\usepackage{amssymb}
\usepackage{amsmath}
\usepackage{natbib}
\usepackage{color}
\usepackage{listings}
\usepackage{graphicx,enumerate}
\usepackage[space]{grffile}
\usepackage{subcaption}
\usepackage{mathtools}
\usepackage{caption}
\usepackage{graphicx,xcolor,colortbl}    
\usepackage{epstopdf,epsfig} 
\usepackage{hyperref}
\usepackage{IEEEtrantools}
\hypersetup{ 
  colorlinks=false,
  pdfborder = {0 0 0.5 [3 0]}
}

\newtheorem{theorem}{Theorem}

\newtheorem{remark}[theorem]{Remark}

\numberwithin{equation}{section} 

\def\Q{{\mathbb Q}}        
\def\R{{\mathbb R}}        
\def\P{{\mathbb P}}        
\def\E{{\mathbb E}}        
 
\def\1{{\mathbf 1}}        
\def\F{{\mathcal F}}        

\def\L{{\mathcal L} \,}

\def\setT{{\mathcal T}}

\def\Vk1{{V^{(i-1)}}}

\def\Hk1{{H^{(i-1)}}}

\def\pk1{{p^{(i-1)}}}

\def\vs{{\varsigma}}

\def\VV{{\mathcal V}}
\def\JJ{{\mathcal J}}
\def\UU{{\mathcal U}}
\def\KK{{\mathcal K}}
\def\PP{{\mathcal P}}
\def\AA{{\mathcal A}}
\def\BB{{\mathcal B}}

\usepackage{multirow}
\usepackage[left=1in,right=1in,top=1.1in,bottom=1.1in]{geometry}

\begin{document}
\title{Trading VIX Futures under Mean Reversion with Regime Switching\thanks{The author would like to thank Professor Tim Leung for many helpful and insightful comments and suggestions.}}
\author{Jiao Li\thanks{APAM Department, Columbia University, New York, NY 10027; email:\,\mbox{jl4170@columbia.edu}.} }
\date{\today}
\maketitle
\begin{abstract}
This paper studies the optimal VIX futures trading problems under a regime-switching model. We consider the VIX as mean reversion dynamics with dependence on the regime that switches among a finite number of states. For the trading strategies, we analyze the timings and sequences of the investor's market participation, which leads to several corresponding coupled system of variational inequalities. The numerical approach is developed to solve these optimal double stopping problems by using projected-successive-over-relaxation (PSOR) method with Crank–-Nicolson scheme. We illustrate the optimal boundaries via numerical examples of two-state Markov chain model. In particular, we examine the impacts of transaction costs and regime-switching timings on the VIX futures trading strategies.

\end{abstract}
\vspace{10pt}

 \newpage
\section{Introduction\label{sect-intro}}
The Chicago Board Options Exchange (CBOE) Volatility Index, commonly known as the  VIX,  is widely used measure of the market volatility. It was introduced   by the CBOE back in 1993, and its calcualtion is  based on the implied volatility of S\&P 500 index options.  Empirically,  the  VIX is shown to be negatively correlated to the market index. One   broadly accepted explanation is that investors buy put  options on S\&P500 for protection against market turmoil. This  increases the option prices and implied volatilities, and thus,  the value of the VIX. As a result,  the VIX is also called the {investor fear gauge} (see \cite{whaley2000investor}).

While VIX itself is not traded, investors can gain exposure to the index  by trading VIX futures.  VIX futures, which are cash settled on the VIX index level, were first traded on  the CBOE Futures Exchange in  2004, and since then the market has been growing constantly. Recent market data illustrate the liquidity and popularity of the VIX futures market. During   2015, the average daily volume (ADV) in the VIX futures was over 200,000 contracts, with  51.6 million VIX futures contracts traded in total.

Each VIX futures comes with a fixed term, ranging from  1 to 9 months, but  futures holders do not need to  keep the position through the expiration date, and can choose when to  close out the position. Furthermore, before  entering the market, the investor can opt to start a  long or short position, followed by  closing  it at later time before expiration. 
 
In this paper, we investigate the VIX futures trading under Cox-Ingersoll-Ross (CIR) model with regime switching. The CIR dynamics is well studied in many empirical studies (see, e.g. \cite{Grunbichler1996985}, \cite{wangVIX2011}). \cite{futures_zhang}  analyze the empirical validity of the CIR model by first estimating its parameters from VIX historical data and provide a futures pricing formula. Later \cite{dotsis2007empirical} add jumps to a CIR diffusion. However, studies of   \cite{futuresVIX} and \cite{futures_SircarAndrew} show that VIX implied volatility data can be better reproduced by incorporating regime switching. \cite{LeungLiLiZheng2015}  display two characteristically different term structures observed in the VIX futures market. These markedly different regimes offer a very interesting testing ground for analyzing the performance of different models for volatility derivatives. It is necessary to allow the key parameters of the VIX to respond to the general market movements. The regime-switching model is one of such formulations, where the model parameters depend on the market regime that switches among a finite number of states. The market regime could reflect the state of the market, the general mood of investors, and other economic factors. \cite{elliott2008pde} evaluate the risk measures for derivatives via a partial differential equation (PDE) approach when the underlying asset dynamics are associated with regime switching. The applications on regime switching for derivatives pricing and optimal stopping problems have been well studied in the literature (see, e.g. \cite{guo2001explicit}, \cite{Buffington2002} and \cite{le2010finite}).
 
Moreover, we introduce the investor's optimal strategies to participate the trading.  In the first strategy, an investor is expected to establish the long position when the price is sufficiently  low, and then  exits when the price is high. The opposite is expected  for the second strategy. In both cases, the presence of transaction costs expands the waiting region, indicating the investor's desire for better prices. In addition, the waiting region expands drastically near expiry since transaction costs discourage entry when futures is very close to maturity. Finally, the main feature of our trading problem approach is to combine these two related problems and analyze the optimal strategy when an investor has the freedom to choose between either a \emph{long-short} or a \emph{short-long} position. Among our results, we find that when the investor has the right to choose, she delays market entry to wait for better prices compared to the individual standalone problems.

In an earlier study,  \cite{futuresBS} propose the optimal strategies for stock index futures with position limits, where the Brownian bridge process is applied for the the basis dynamics. \cite{futuresDaiKwok} extend the optimal strategies with the additional flexibility by allowing the investor to switch between long position and short position directly. In this paper,  we extend the model introduced in \cite{LeungLiLiZheng2015} 
by incorporating regime switching into the CIR model for the VIX. We provide a link between the futures pricing problem under the risk-neutral measure and the trading problem conducted under the historical measure. In contrast to \cite{futuresBS} and \cite{futuresDaiKwok}, we do not a priori assume the existence of a stochastic basis that may or may not be consistent with the futures prices, but consider the long and short strategies that take advantage of the temporal price difference of futures. Similar trading strategies have been studied by \cite{LeungLi2014OU}, \cite{LeungLiWang2014XOU}, \cite{LeungLiWang2014CIR}, \cite{meanreversionbook2016}, \cite{LeungShirai} and \cite{stepanek2015comparison}, among others. Moreover, the strategy studied herein can be automated. For a comprehensive study on algorithmic trading, we refer the reader to \cite{HFTbook}. The ideas can be applied to other derivatives, such as swaps (see \cite{LeungYamazakiQF} and \cite{LeungLiu2012}). 

 

With regime switching, the optimal trading strategies are determined from a system of $m$ variational inequalities, where $m$ is the number of possible regimes of the market. We apply an implicit-explicit finite difference method with projected-successive-over-relaxation (PSOR) to solve for the optimal trading boundaries under all regimes. While there are a number of studies on optimal stopping problems with regime switching as mentioned above, there are very few that discuss the numerical methods and solutions. In a related study, \cite{khaliq2009new} develop a penalty method method for pricing the regime-switching American option.

The rest of the paper is organized as follows.  In section \ref{sect-futtradingregime}, we formulate the optimal stopping and trading strategies with regime switching. In section \ref{numerical}, we develop an implicit scheme based on PSOR method. In Section \ref{explanation-results}, we present the numerical results for optimal trading problems and provide financial interpretations.



%
%
%

\section{Optimal Timing to Trade Futures with  Regime Switching\label{sect-futtradingregime}}

We fix a probability space $(\Omega, \F, \P)$, where $\P$ is the historical measure. Let $\xi$ be a continuous-time irreducible finite-state Markov chain with state space $E=\{1, 2, \ldots, m\}$. The generator matrix of $\xi$ is denoted by $Q$, which has constant entries $q_{ij}$ for $i,j\in E$, such that $q_{ij}\geq 0$ for $i\neq j$ and $\sum_{j\in E} q_{ij} = 0$ for each $i\in E$. This Markov chain represents the changing regime of the financial market, and it influences the dynamics of the index.

Let $B$ be a standard Brownian motion defined on $(\Omega, \F, \P)$ and assume it is independent of $\xi$. The VIX, denoted by $S$, is assumed to follow the CIR process regime switching:
\begin{align}\label{CIRP_RS}
dS_t = {\mu}(\xi _t)( {\theta}(\xi _t)-S_t)dt+\sigma(\xi _t) \sqrt{S_t} dB_t,
\end{align}
where, for each $i\in E$, the coefficients ${\mu}(i)$, ${\theta}(i)$ and $\sigma(i)$ are known constants, with ${\mu}(i)$, ${\theta}(i)$, $\sigma(i) > 0$. Note that ${\mu}(i)$, ${\theta}(i)$ and $\sigma(i)$ depend on the Markov chain, representing the mean reversion rate, the long-run mean and the volatility of the VIX at regime $i$.

To price futures, we assume a re-parametrized CIR model for  the risk-neutral VIX dynamics. Due to the additional uncertainty described by regime switching, we note that introducing a Markov chain results in an incomplete market. Hence the equivalent martingale measure is not unique. \cite{elliott2008pde} employ the regime-switching Esscher transform to determine an equivalent martingale pricing measure. We do not include the argument in our paper and instead assume that the risk-neutral probability space $(\Omega, \F, \Q)$ is given. Thus, under the risk-neutral measure $\Q$, the VIX follows
\begin{align}\label{CIRQ_RS}
dS_t = \tilde{\mu}(\xi _t)(\tilde{\theta}(\xi _t)-S_t)dt + \sigma(\xi _t)\sqrt{S_t} dB_t^{\Q},
\end{align}
where $\tilde{\mu}(i)$, $\tilde{\theta}(i)$, $\sigma(i) > 0$ and $B^{\Q}$ is a  $\Q$-standard Brownian motion which is also independent of Markov chain. For convenience, we may use the subscript notation for these constants, e.g. $\mu _i \equiv {\mu}(i)$, $\theta _i \equiv {\theta}(i)$, and $\sigma _i \equiv {\sigma}(i)$.  In both SDEs,  \eqref{CIRP_RS}  and \eqref{CIRQ_RS}, we require   $2\mu_i \theta_i \ge \sigma_i ^2$ and $2\tilde{\mu}_i \tilde{\theta}_i \ge \sigma_i ^2$ (Feller condition) so that the CIR process stays strictly positive at all times. The two Brownian motions are related by 
\begin{align}
dB_t^{\Q}=d{B}_t+\frac{{\mu (\xi _t)}({\theta (\xi _t)}-S_t)-\tilde{\mu}(\xi _t) (\tilde{\theta} (\xi _t)-S_t)}{\sigma (\xi _t) \sqrt{S_t}}  \,dt,
\end{align}
such as change of measure preserves the CIR model, up to different parameter values across two measures.  

We consider a futures contract written on the $S$ with maturity $T < \infty$ and futures price at time $0 \leq t \leq T$ when $S_t = s$ and regime $\xi _t = i$. The price of a futures contract is given by
\begin{align}\label{fTCIR_RS}
f(t, s, i) = \E^{\Q}\{S_T|S_t = s, \xi_t = i\}, \quad t\le T, 
\end{align}
where $\E ^{\Q}$ is the expectation operator with respect to the risk-neutral measure $\Q$. We can show that  $f_i (t, s) \equiv f(t, s, i)$, $i = 1, \hdots, m$, satisfy the following partial differential equation (see, e.g. \cite{yao2006regime}),
\begin{align}
\begin{cases} 
\begin{split}
\frac{\partial f_i}{\partial t} + \tilde{\mu}_i (\tilde{\theta}_i - s) \frac{\partial f_i}{\partial s} + \frac{\sigma^2 _i s}{2} \frac{\partial ^2 f_i}{\partial s^2} + \sum _{j \neq i} \tilde{q}_{ij} (f_j -f_i) &= 0,   \quad (t,s) \in [0, T) \times \R_+, 
\\ f_i (T,s) &= s,  \quad s \in \R_+,
\end{split}	
\end{cases} \label{fTPDE_RS}				
\end{align}
where $\tilde{q}_{ij}$ is the transition probability under measure $\Q$. We note that (\ref{fTPDE_RS}) involves $m$ interconnected PDEs due to the regime switching introduced in the VIX dynamics. One can efficiently compute the futures prices in all regimes by finite difference methods; see section \ref{numerical} below.
 
\subsection{Optimal Double Stopping Approach}
Let us consider the scenario in which  an investor has a long position in a futures contract with expiration date $T$. With a long position in the futures, the investor can hold it till maturity, but can also close the position  early by taking an opposite position at the prevailing  market price. At maturity, the two opposite positions cancel each other.   This motivates us to investigate the best time to close.

If the  investor selects to close the long position at time $\tau\le \hat{T}$,  then she will receive the market value of the futures on the expiry date, denoted by  $f(\tau,S_\tau, \xi _{\tau})$, minus the transaction cost $c \ge 0$.  To maximize the expected discounted value, evaluated under the investor's historical probability measure $\mathbb{P}$ with a constant subjective discount rate $r\ge 0$,  the investor solves the optimal stopping problem 
 \begin{align}
\VV(t, s, i) = \sup_{\tau \in\setT_{t,\hat{T}}} \E \! \left\{ e^{-r (\tau-t)}(f (\tau, S_{\tau}, \xi _{\tau}) - c) | S_t = s, \xi_t = i \right\}, \label{VV}
\end{align}
where  $\setT_{t,\hat{T}}$  is    the set of all stopping times, with taking values between $t$ and $\hat{T}$, where  $\hat{T} \in (0, T]$ is the trading deadline, which can equal but not exceed the futures' maturity.  Throughout this chapter, we continue to use the notation $\E \{\cdot\}$ to  indicate  the expectation taken under the historical probability measure $\mathbb{P}$.

The value function $\VV(t,s,i)$ represents the expected  liquidation value associated with the  long futures position in $i-$state. Prior to taking the  long position in $f_i$, the investor, with zero position, can select the optimal timing to start the trade, or not to enter at all. This leads us to analyze the timing option  inherent in the trading
problem. Precisely, at time $t\le \hat{T}$, the investor  faces the optimal entry timing problem  
\begin{align}\JJ(t, s, i) =  \sup_{\nu \in \setT_{t, \hat{T}}
}\E \! \left\{e^{-r (\nu-t)} (  \VV(\nu, S_{\nu}, \xi _{\nu})  - (f(\nu, S_{\nu}, \xi _{\nu}) + \hat{c}))^{+} | S_t = s, \xi_t = i  \right\},\label{JJ}
\end{align} 
where $\hat{c} \ge 0$ is the transaction cost, which may differ from $c$. In other words, the investor seeks
to maximize the expected difference between the value function
$\VV(\nu, S_\nu, \xi_\nu)$ associated with the long position and the prevailing   futures price $f(\nu, S_{\nu}, \xi_{\nu})$. The value function $\JJ(\nu, S_\nu, \xi_\nu)$
represents the maximum expected value of the  trading opportunity embedded in the futures.  We refer this ``long to open, short to close" strategy as the \emph{long-short} strategy\index{long-short strategy}.

Alternatively, an investor may well  choose to short a futures contract with the speculation that the futures price will fall.  By taking a short futures position, the  investor can either close it out later by establishing a long position, or hold it until the expiry which will result in a cash settlement. Given  an investor  who has  a unit  short position in the  futures contract, the objective is to minimize the expected discounted cost to close out this position at/before maturity.  The optimal timing strategy  is determined from 
\begin{align}
\UU(t, s, i) =  \inf_{\tau \in \setT_{t,\hat{T}}
}\E \! \left\{e^{-r (\tau-t)}(f(\tau, S_{\tau}, \xi _{\tau}) + \hat{c}) | S_t = s, \xi_t = i \right\}. \label{UU}
\end{align} 
If the investor  begins with a zero position, then she can decide when to enter the market by solving
\begin{align}
\KK(t, s, i) =  \sup_{\nu \in \setT_{t,\hat{T}}
}\E \!\left\{e^{-r (\nu-t)} ((f(\nu, S_{\nu}, \xi _{\nu}) - c) - \UU(\nu, S_{\nu}, \xi _{\nu}))^{+} | S_t = s, \xi_t = i \right\}.\label{KK}
\end{align}
We call this  ``short to open, long  to close" strategy as the \emph{short-long} strategy\index{short-long strategy}. 

When an investor contemplates entering the market, she can either long or short first. Therefore, on top of the timing option, the investor has an additional choice between the long-short and short-long strategies. Hence, the investor solves the market entry timing problem:
 \begin{align}
\PP(t, s, i) =  \sup_{\vs \in \setT_{t,\hat{T}}
}\E \!\left\{e^{-r (\vs-t)} {\max}\{\mathcal{A}(\vs, S_\vs, \xi_{\vs}), \mathcal{B}(\vs, S_\vs, \xi_{\vs})\} | S_t = s, \xi_t = i \right\},\label{PP}
\end{align}
with two alternative rewards upon entry defined by
\begin{align*}
\mathcal{A}(\vs, S_{\vs}, \xi_{\vs}) &:= (\VV(\vs, S_{\vs}, \xi_{\vs})  - (f(\vs, S_{\vs}, \xi_{\vs}) + \hat{c}))^{+} \quad \text{(long-short)},\\
\mathcal{B}(\vs,S_{\vs}, \xi_{\vs}) &:= ((f(\vs, S_{\vs} , \xi_{\vs}) - c) -\UU(\vs, S_{\vs},  \xi_{\vs}) )^{+} \quad \text {(short-long)}.
\end{align*} 
Accordingly, the corresponding inputs associated with the optimal stopping problem in \eqref{PP} is given by taking the maximum between the above two rewards.

\subsection{Variational Inequalities \& Optimal Trading Strategies}\label{sect-VIRS}
Given the CIR dynamics of the VIX, the value functions defined in \eqref{VV} -- \eqref{PP} all satisfy the same governing differential equation in their respective continuation regions and their values equal to the corresponding rewards in their exercise regions. In order to solve for the optimal trading strategies, we need to study the coupled systems of variational inequalities respectively. To this end, we first define the operators:
 \begin{align}
 \L_{i}\{\cdot\}&:= -r \cdot + \frac{\partial\cdot}{\partial t} + \mu_i( \theta_i  - s)\frac{\partial\cdot}{\partial s} + \frac{\sigma_i^2s}{2}\frac{\partial^2 \cdot}{\partial s^2} + \sum _{j \neq i} q_{ij} (\cdot _j - \cdot _i) \label{lCIR},
\end{align}
corresponding to CIR model. For convenience, we adopt the subscript notation for these value functions, e.g. $\VV_i(t,s) \equiv \VV(t,s,i)$, $\JJ_i(t,s) \equiv \JJ(t,s,i)$, $\UU_i(t,s) \equiv \UU(t,s,i)$, $\KK_i(t,s) \equiv \KK(t,s,i)$ and $\PP_i(t,s) \equiv \PP(t,s,i)$.

The optimal exit and entry problems  $\JJ_i$ and $\VV_i$ associated with the  \emph{long-short} strategy    are solved from  the following pair of variational inequalities:
\begin{align}
\textrm{max}\left\{\,\L_{i} \VV_i(t,s)\,, \,(f_i(t, s) -  c) - \VV_i(t,s)\,\right\} &= 0,\label{VIV_RS}\\
\textrm{max}\left\{\,\L_i \JJ _i (t,s)\, ,\, (\VV_i(t,s)- (f_i(t, s)+\hat{c}))^{+} - \JJ_i(t,s) \, \right\} &=0, \label{VIJ_RS}
\end{align}
for $(t,s) \in [0,\hat{T}]\times  \mathbb{R_+}$. Similarly, the reverse \textit{short-long} strategy can be determined by numerically solving the variational inequalities satisfied by $\UU_i$ and $\KK_i$:
\begin{align}
\textrm{min}\left\{\,\L_i \UU_i(t,s)\,,\,  (f_i(t, s) + \hat{c}) -\UU_i(t,s) \,\right\} &= 0,\label{VIU_RS}\\
\textrm{max}\left\{\,\L_i \KK_i(t,s)\, , \,((f_i(t, s) - c) - \UU_i(t, s))^{+} - \KK_i(t,s) \, \right\} &=0. \label{VIK_RS}
\end{align}
To determine the optimal timing to enter the futures market, we  solve the variational inequality 
 \begin{align}
\textrm{max}\left\{\,\L_i \PP_i(t,s)\,,\, \textrm{max}\{\mathcal{A}_i(t,s), \mathcal{B}_i(t,s)\} - \PP_i(t,s)\,  \right\} &=0. \label{VIP_RS}
\end{align}
In other words, we have to solve the  variational inequality \eqref{VIV_RS} -- \eqref{VIK_RS}, and then use the solution as inputs to the variational inequality \eqref{VIP_RS}.

\section{Numerical Implementation with Regime Switching\label{numerical}}
The numerical solution of the system of variational inequalities can be obtained by applying a finite-difference scheme in all regimes with the use of the projected-successive-over-relaxation (PSOR) method.\footnote{We refer to Chapter 9 of \cite{wilmottbook1995} for a detailed discussion on the projected SOR method.} The solution of the resulting equations for value functions are solved by the successive over relaxation (SOR) method. In each SOR iterative step in finding the numerical approximation of the value functions, we simply take the maximum value between the approximated function value and compensated futures price. The futures price can be pre-computed by \eqref{fTPDE_RS} conveniently. Similar numerical schemes with regime switching can be found in \cite{Leung_regime2010}. More details of our numerical scheme are described in the following. 

First, we define the generic differential operator
\begin{align}
 \L_i\{\cdot \}:= -r \cdot + \frac{\partial\cdot}{\partial t} + \varphi _i (s) \frac{\partial\cdot}{\partial s} + \frac{\sigma _i^2(s)}{2}\frac{\partial^2 \cdot}{\partial s^2} + \sum _{j \neq i} q_{ij} (\cdot _j - \cdot _i), \label{Loperator}
\end{align}
then the variational inequalities \eqref{VIV_RS} -- \eqref{VIP_RS} admit the same form as the following variational inequality problem:
\begin{align}
\begin{cases} 
\begin{split}
\L_i g_i(t,s) \leq 0, \enspace g_i(t,s) & \geq h_i (t,s),  \quad (t,s) \in [0,\hat{T}) \times \R_+,  \\ 
\\ (\L_i g_i(t,s)) (h_i (t,s) - g_i(t,s)) &= 0,  \quad (t,s) \in [0,\hat{T}) \times \R_+,\\
\\ g_i (\hat{T},s) &= h_i (\hat{T},s),  \quad s \in \R_+.
\end{split}	
\end{cases} \label{VIg}				
\end{align}
Here, $g_i(t,s)$   represents the value functions $\VV_i(t,s)$, $\JJ_i (t,s)$, $-\UU_i (t,s)$, $\KK_i (t,s)$, or $\PP_i(t,s)$. The function $h_i(t,s)$ represents $f_i(t,s) - c$, $(\VV_i(t,s) - (f_i(t,s) + \hat{c}))^+$, $-(f_i(t,s) + \hat{c})$, $(f_i(t,s) - c) - \UU_i(t,s))^+$, or $\max\{ \AA_i(t,s), \BB_i(t,s) \}$. The futures price $f_i(t,s)$, with $\hat{T} \le T$, is given by \eqref{fTCIR_RS}.

We now consider the discretization of the partial differential equation $ \L _i g_i(t,s) =0$, over an uniform grid with discretizations in   time ($\delta t = \frac{\hat{T}}{N}$), and space    ($\delta s = \frac{S{\max}}{M}$).  Applying  the Crank-Nicolson method for $s$-derivatives and backward difference for $t$-derivatives on the resulting equation leads the finite difference equation:
\begin{multline}
- \frac{r}{2} (g_i ^{m,n} + g_i ^{m,n-1}) + \frac{g_i ^{m,n}-g_i ^{m,n-1}}{\delta t}+\frac{\varphi _i ^m}{2} (\frac{g_i ^{m+1,n}-g_i ^{m-1,n}}{2\delta s} + \frac{g_i ^{m+1,n-1}-g_i ^{m-1,n-1}}{2\delta s})\\
+\frac{(\sigma _i ^m) ^2}{2} (\frac{g_i ^{m+1,n}-2g_i ^{m,n}+g_i ^{m-1,n}}{2 (\delta s)^2} + \frac{g_i ^{m+1,n-1}-2g_i ^{m,n-1}+g_i ^{m-1,n-1}}{2 (\delta s)^2}) \\
+ \frac{q_{ii}}{2}(g_i ^{m,n} + g_i ^{m,n-1}) + \sum _{j \neq i} \frac{q_{ij}}{2} (g_j ^{m,n} + g_j ^{m,n-1})=0,
\label{FDgrid_RS}
\end{multline}
where $q_{ii} = - \sum _{j \neq i} q_{ij}$ is used. For convenience, we may use the subscript notation for these constants, e.g. $g_i ^{m,n} \equiv g_i(n \delta t, m \delta s )$, $h_i ^{m,n} \equiv h_i (n \delta t, m \delta s )$, $\varphi ^m _i = \varphi_i (m \delta s)$ and $\sigma^m _i = \sigma_i (m \delta s)$. We implement by explicitly treating the regime coupling terms. Replace $g_j ^{m,n-1}$ with $g_j ^{m,n}$ for $j \neq i$ and obtain
\begin{multline}
-\alpha_i ^m g_i ^{m-1,n-1} + (1-\beta_i ^m) g_i ^{m,n-1} - \gamma_i ^m g_i ^{m+1, n-1}\\
=\alpha_i ^m g_i ^{m-1,n} + (1+\beta_i ^m) g_i ^{m,n} + \gamma_i ^m g_{m+1,n} + \delta t \sum _{j \neq i} q_{ij} g_j ^{m,n}, \label{scheme_RS}
\end{multline}
where   
\begin{align}
\begin{cases} 
\begin{split}
\alpha_i ^m &= \frac{\delta t}{4 \delta s}\big( \frac{(\sigma^m _i)^2 }{\delta s} - \varphi _i ^m \big),  \\ 
\\ \beta_i ^m &= -\frac{\delta t}{2} \big((r-q_{ii}) + \frac{(\sigma_i ^m) ^2}{(\delta s)^2}\big),\\
\\ \gamma_i ^m &= \frac{\delta t}{4 \delta s}\big( \frac{(\sigma_i ^m) ^2  }{\delta s} + \varphi ^m _i \big),
\end{split}	
\end{cases} \label{coef_RS}		
\end{align}
%
for $i, j \in E$, $m=1,2,...,M-1$ and $n=1,2,...,N-1$. The system to be solved  backward in time is 
\begin{align}
\mathbf{M_i ^1 g_i ^{n-1}=r_i ^n},\label{M1g}
\end{align}
where the right-hand side is 
\begin{align}
\mathbf{r_i ^n=M_i ^2 g_i^{n}} +\delta t \sum _{j \neq i} q_{ij} \mathbf{g_j ^n}+\alpha^1 _i \begin{bmatrix} g_i ^{0,n-1}+g_i ^{0,n} \\ 0 \\ \vdots \\0 \end{bmatrix} + \gamma_i ^{M-1} \begin{bmatrix}  0 \\ \vdots \\0 \\ g_i ^{M,n-1}+g_i ^{M,n} \end{bmatrix},
\end{align}
and
\begin{align}
\mathbf{M_i ^1} &= \left[ \begin{array}{cccccc}
1- \beta _i ^1 & -\gamma_i ^1 & & & \\
-\alpha _i ^2 & 1- \beta _i ^2 & -\gamma_i ^2 & & \\
& -\alpha _i ^3 & 1- \beta _i ^3 & -\gamma_i ^3 & \\
& & \ddots & \ddots & \ddots \\
& & &- \alpha_i ^{M-2} & 1- \beta _i ^{M-2} & -\gamma_i ^{M-2} \\
& & & & - \alpha_i ^{M-1} & 1- \beta _i ^{M-1} \end{array} \right],\\
\mathbf{M_i ^2} &= \left[ \begin{array}{cccccc}
1+ \beta _i ^1 & \gamma_i ^1 & & & \\
\alpha _i ^2 & 1+ \beta _i ^2 & \gamma_i ^2 & & \\
& \alpha _i ^3 & 1+ \beta _i ^3 & \gamma_i ^3 & \\
& & \ddots & \ddots & \ddots \\
& & & \alpha_i ^{M-2} & 1+ \beta _i ^{M-2} & \gamma_i ^{M-2} \\
& & & & \alpha_i ^{M-1} & 1+ \beta _i ^{M-1} \end{array} \right],\\
\mathbf{g_i ^n} &=\begin{bmatrix} g_i ^{1,n}, g_i ^{2,n}, \hdots , g_i ^{M-1,n} \end{bmatrix} ^T.
\end{align}
We note that the dimension of $\mathbf{M_i ^1}$ is independent of $m$, which is the number of regimes. Thus the scheme \eqref{M1g} can be computed in parallel. 

\begin{remark} 
The futures price function \eqref{fTPDE_RS} can be computed via solving the linear system \eqref{M1g} by replacing $g_i ^{m,n}$ with $f_i ^{m,n}$, where $f_i ^{m,n} \equiv f(t,s,i)$, and setting $r = 0$.
\end{remark}

Since the investor can establish her position at anytime before the expiry, the value functions $g_i (t, s)$ must satisfy the constraint
\begin{align}
g_i(t,s) \geq h_i (t,s),  \quad s \geq 0,\quad  0 \leq t \leq \hat{T}, \quad i \in E ,
\end{align}
where the discrete scheme can be written as 
\begin{align}
g_i ^{m, n} \geq h_i ^{m, n}, \quad 0 \leq m \leq M, \quad  0 \leq n \leq M, \quad i \in E.
\end{align}
Hence, at each time step $n \in \left\{1, 2, \hdots, N-1\right\}$,  we need to solve 
\begin{align}
\begin{cases} 
\begin{split}
\mathbf{M_i ^1 g_i ^{n-1}} & \geq \mathbf{r_i ^n}, \\
\\ \mathbf{g_i ^{n-1}} & \geq \boldsymbol{h_i ^{n-1}},   \\ 
\\ (\mathbf{M_i ^1 g_i ^{n-1}} -\mathbf{r_i ^n})^T (\boldsymbol{h _i ^{n-1}} - \mathbf{g_i ^{n-1}}) &= 0.  
\end{split}	
\end{cases} 			
\end{align}
To guarantee the constraint, our algorithm enforces the constraint explicitly as follows
\begin{align}
g^{m,n-1}_{i,new}=\max \big\{g^ {m,n-1} _{i,old},h^{m,n-1} _i\big\}.
\label{iterative}
\end{align}
The   projected SOR method is used to solve the linear system. Notice that the constraint is enforced at the same time as the iterate $g^{m,n-1}_{i,(k+1)}$ is calculated; the effect of the constraint is immediately felt in the calculation of $g^{m+1,n-1}_{i,(k+1)}$, $g^{m+2,n-1}_{i,(k+1)}$, etc. Thus, at each time step $n$, the PSOR algorithm is to iterate (on $k$) the equations
\begin{align} 
\begin{split}
g^{1,n-1}_{i,(k+1)} &= \max \big\{h_i ^{1,n-1} \,,\, g^{1,n-1}_{i, (k)} + \frac{\omega}{1-\beta_i ^1} [r_i ^{1,n}-(1-\beta_i^1) g^{1,n-1}_{i, (k)}+\gamma_i ^1 g^{2,n-1}_{i, (k)}] \big\},\\
g^{2,n-1}_{i, (k+1)} &= \max \big\{h_i ^{2,n-1} \,,\, g^{2,n-1}_{i, (k)} + \frac{\omega}{1-\beta_i^2} [r_i ^{2,n}+\alpha_i^2 g^{1,n-1}_{i,(k+1)}-(1-\beta_i^2) g^{2,n-1}_{i,(k)}+\gamma_i ^2 g^{3,n-1}_{i,(k)}] \big\},\\
\vdots\\
g^{M-1,n-1}_{i,(k+1)} &= \max \big\{h_i ^{M-1,n-1} \,,\, g^{M-1,n-1}_{i,(k)} \\
&+ \frac{\omega}{1-\beta_i ^{M-1}} [r_i^{M-1,n}+\alpha_i ^{M-1} g^{M-2,n-1}_{i,(k+1)}-(1-\beta_i ^{M-1}) g^{M-1,n-1}_{i,(k)}] \big\},
\end{split}
\label{PSOR}					
\end{align}
where $k$ is the iteration counter and $\omega$ is the overrelaxation parameter.
The iterative scheme starts from an initial point $\mathbf{g}^n _{i,(0)}$ and proceeds until a convergence criterion is met, such as
$|| \mathbf{g^{n-1} _{i, \mathit{(k+\mathrm{1})}}} - \mathbf{g^{n-1} _{i,\mathit{(k)}}} || < \epsilon ,$ where $\epsilon$ is a tolerance parameter.  The  optimal boundary $S_f(t)$ can be identified by  locating the boundary that separates the  regions where  $g_i(t,s)=h_i(t,s)$, or  $g_i(t,s) > h_i(t,s)$. 

\begin{figure}[h]
   \centering
\begin{subfigure}[b]{0.48\textwidth}
    \includegraphics[width=\textwidth]{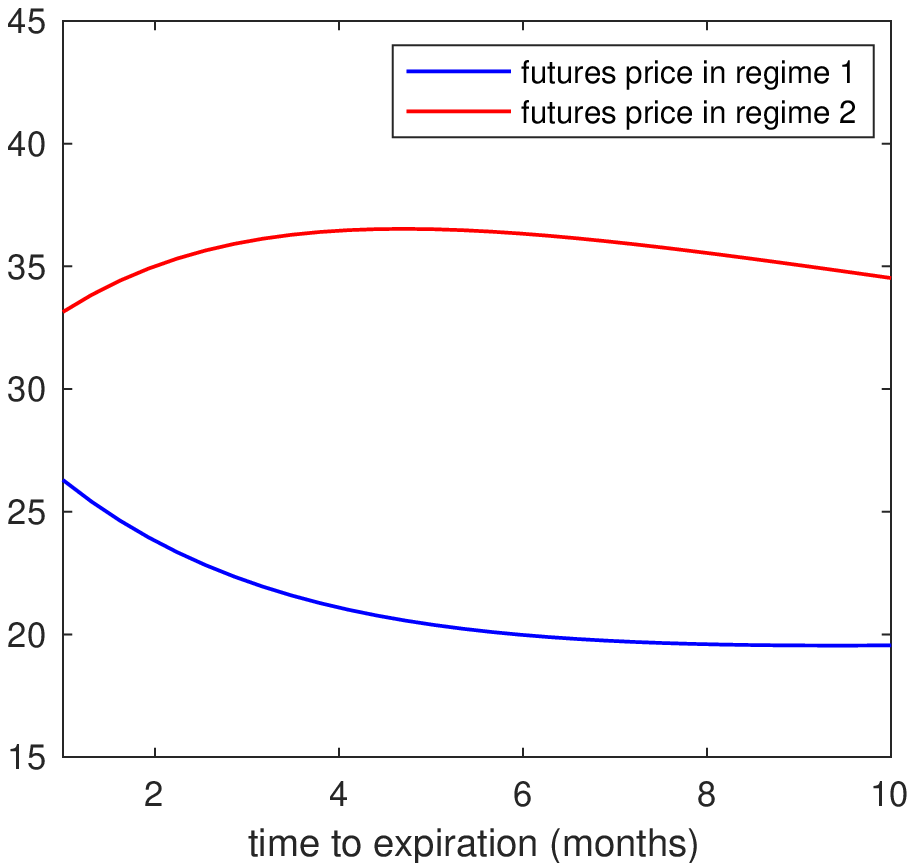}
\end{subfigure}
 \caption{The futures prices in 2 regimes. Parameters: $S_0 = 30,  \sigma_1=5.33, \sigma_2=6.42,  \tilde{\theta}_1=18.16, \tilde{\theta}_2=40.36, \tilde{\mu}_1=4.55, \tilde{\mu}_2=4.59, q_{12}=-q_{11}= 0.1, q_{21} = - q_{22}=0.5.$}\label{Futures_Price}
\end{figure}

\clearpage

\section{Optimal Trading Strategies\label{explanation-results}}
\begin{figure}[h]
    \centering
\begin{tabular}[h]{@{}p{0.35\hsize}@{}p{0.35\hsize}@{}}
\begin{subfigure}{\hsize}\centering
\includegraphics[width=2.4in]{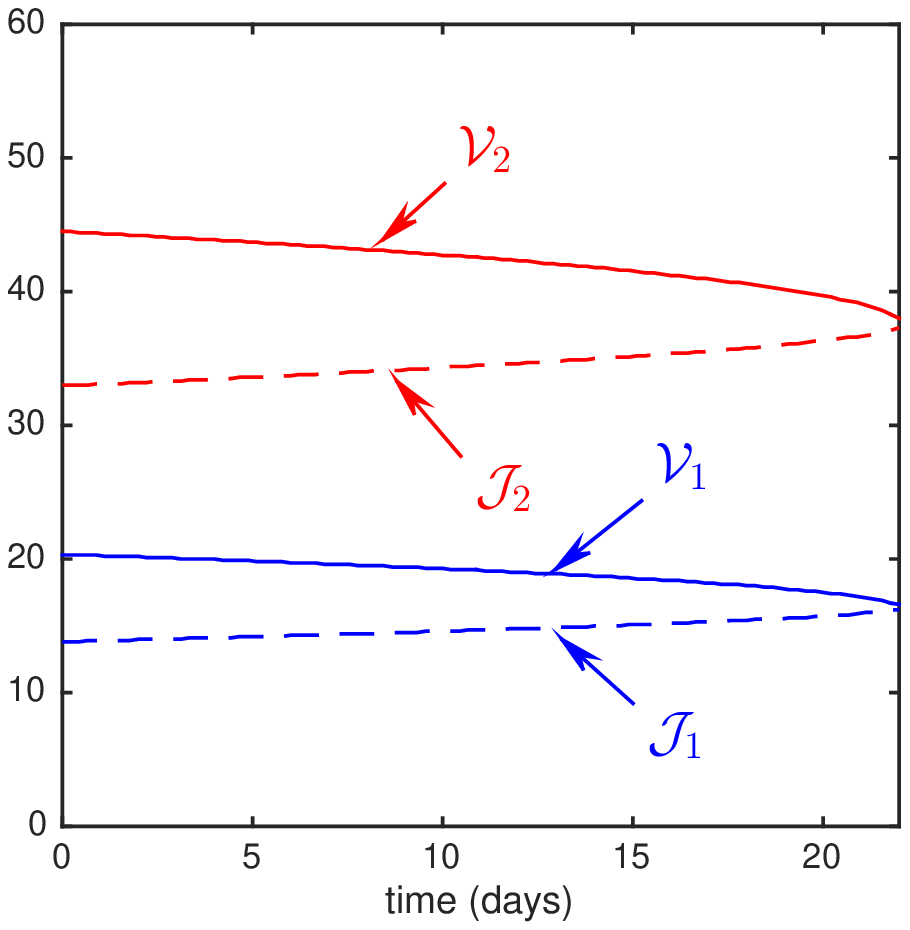}
\caption{}\label{CIRVVJJ0}
\end{subfigure}                     \\
\begin{subfigure}{\hsize}\centering
\includegraphics[width=2.4in]{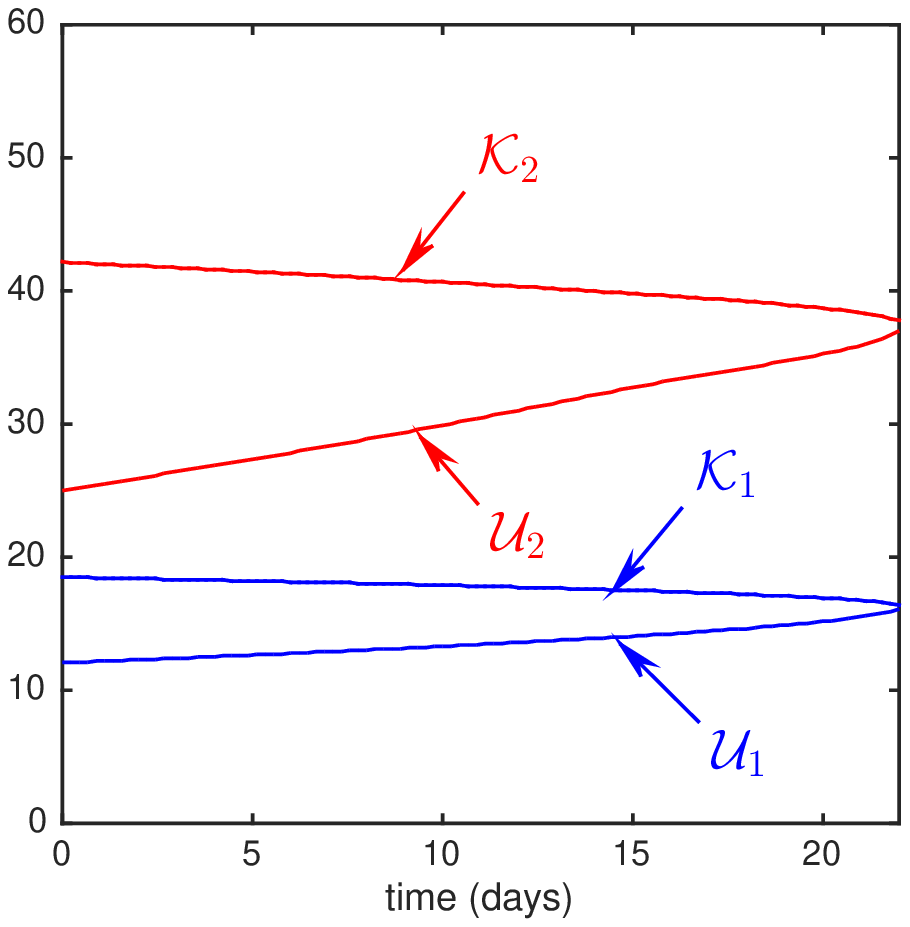}
\caption{}\label{CIRKKUU0}
\end{subfigure}      \\
\begin{subfigure}{\hsize}\centering
 \includegraphics[width=2.4in]{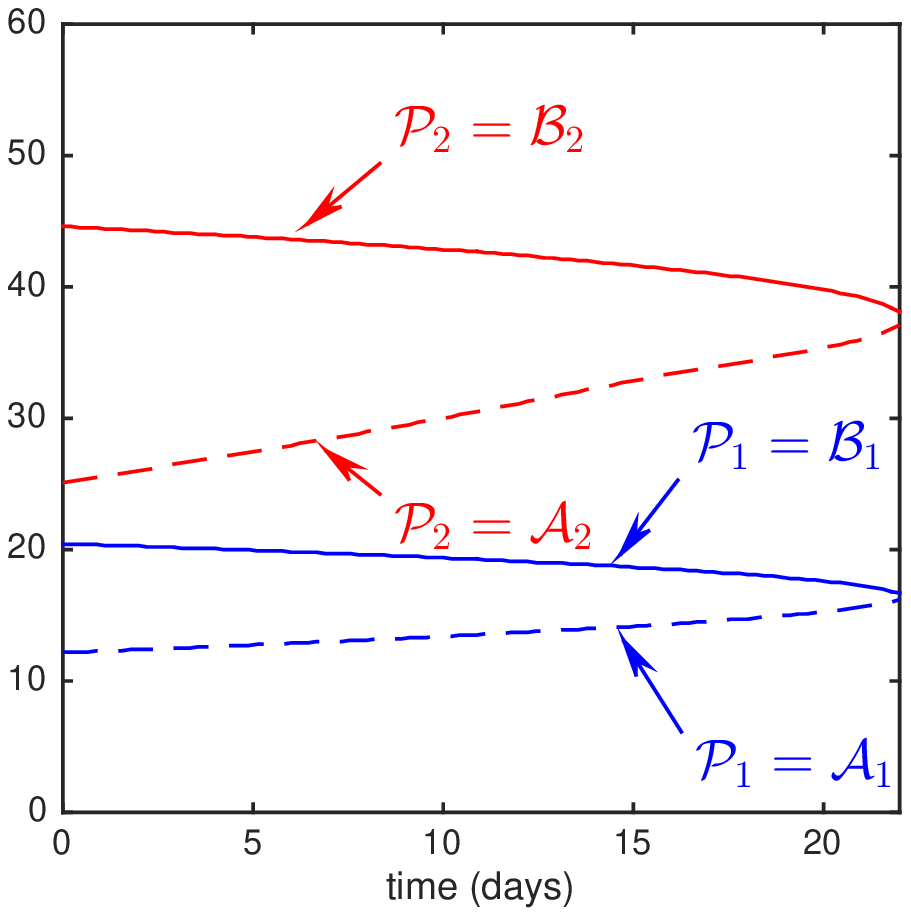}
 \caption{}\label{CIRPP0}
\end{subfigure}
\end{tabular}
\begin{tabular}[h]{@{}p{0.35\hsize}@{}p{0.35\hsize}@{}}
\begin{subfigure}{\hsize}\centering
\includegraphics[width=2.4in]{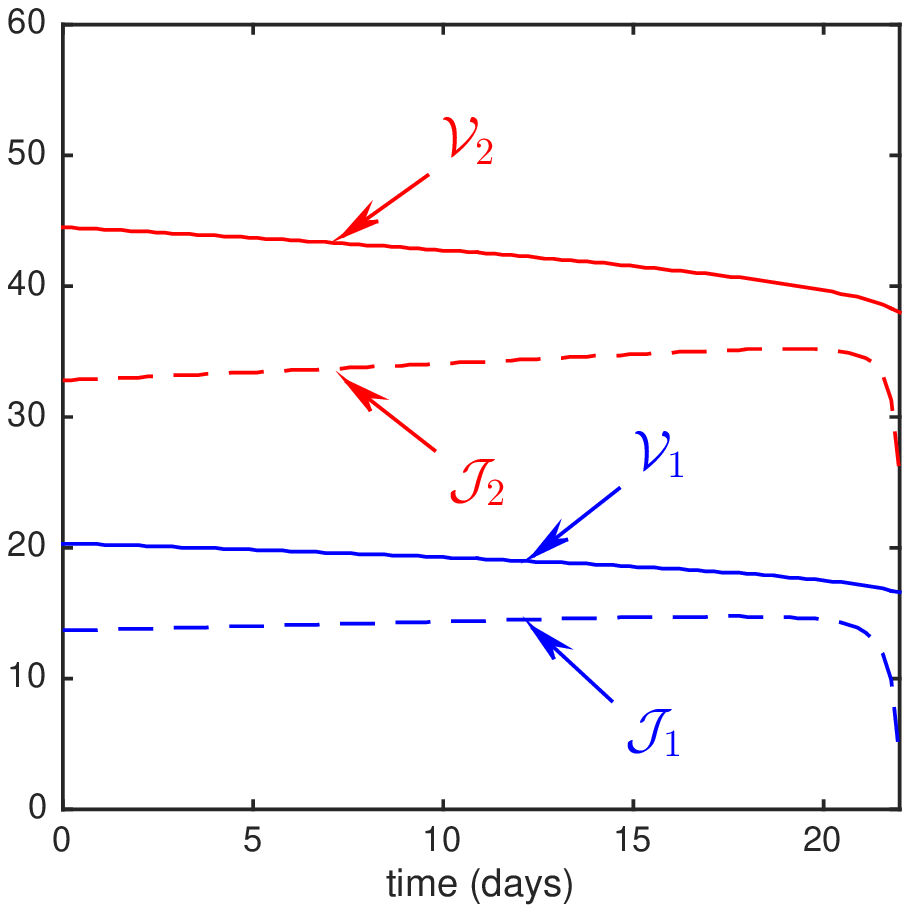}
\caption{}\label{CIRVVJJ1}
\end{subfigure}                     \\
\begin{subfigure}{\hsize}\centering
\includegraphics[width=2.4in]{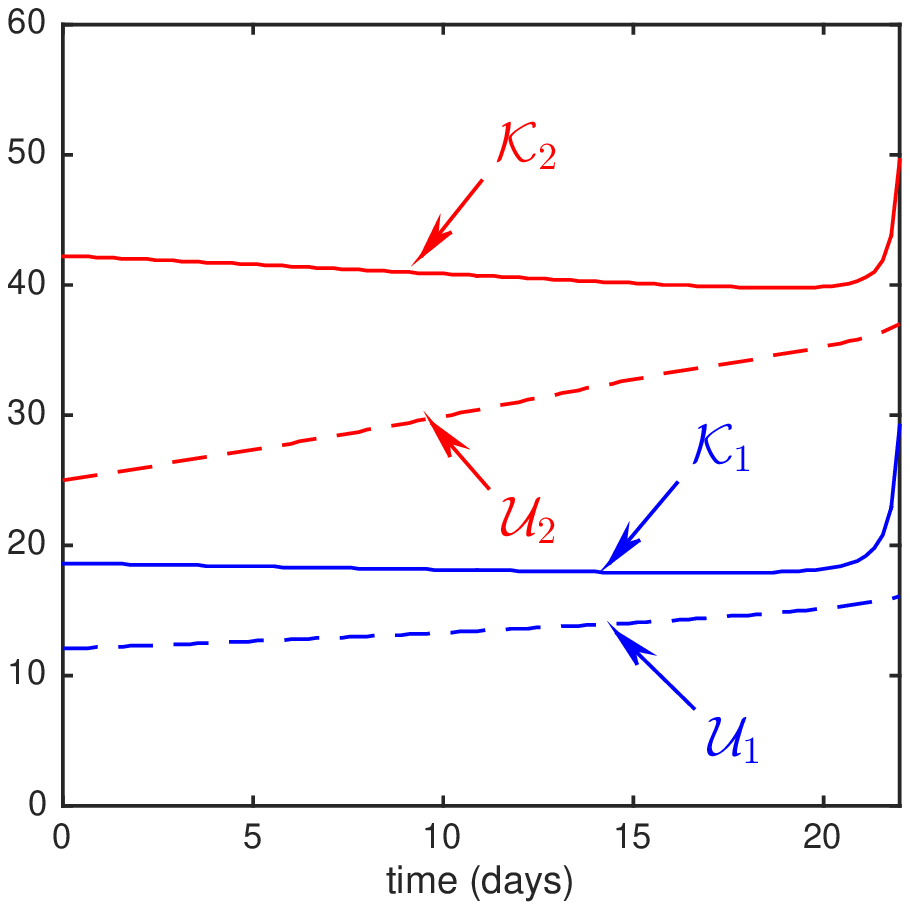}
\caption{}\label{CIRKKUU1}
\end{subfigure}      \\
\begin{subfigure}{\hsize}\centering
 \includegraphics[width=2.4in]{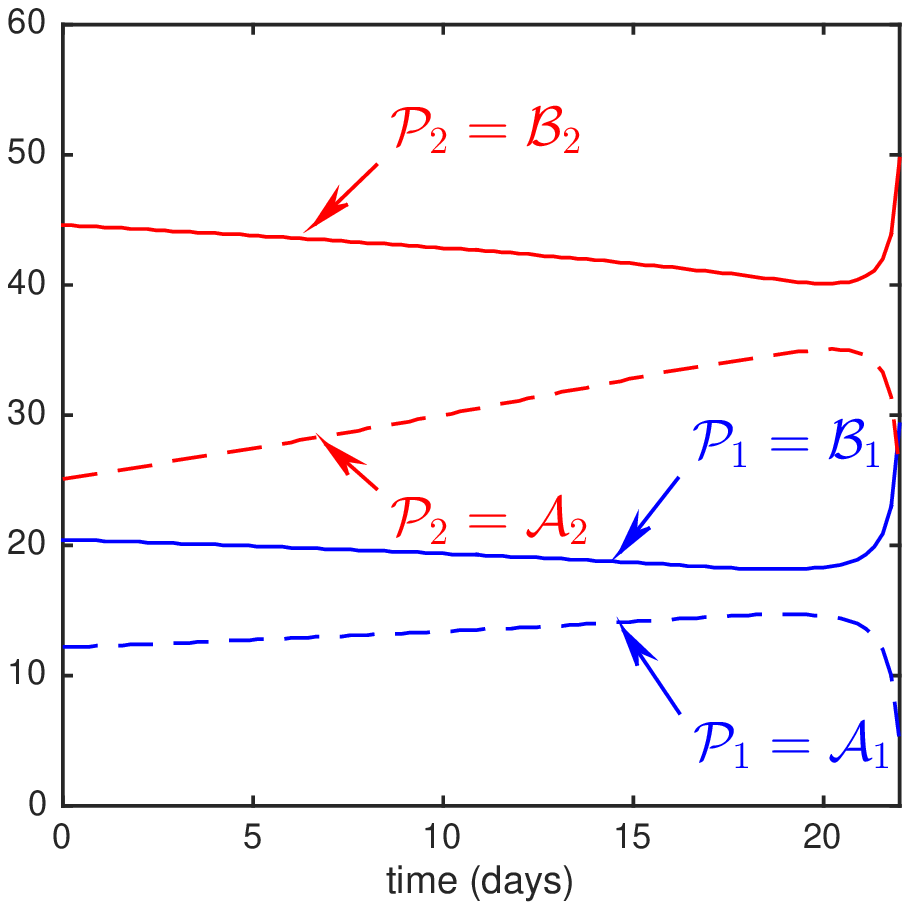}
 \caption{}\label{CIRPP1}
\end{subfigure}
\end{tabular}
\caption{Optimal long-short boundaries for futures trading with 2-state regime-switching model. Left panel: $c=\hat{c}=0$. Right panel: $c=\hat{c}=0.01$. Common parameters: $\hat{T}=\frac{22}{252}$ , $T=\frac{66}{252}$, $ r=0.05, \sigma_1=5.33, \sigma_2=6.42, \theta_1= 17.58, \theta_2= 39.5, \tilde{\theta}_1=18.16, \tilde{\theta}_2=40.36, \mu _1=8.57, \mu _2=9, \tilde{\mu}_1=4.55, \tilde{\mu}_2=4.59, q_{12}=-q_{11}= 0.1, q_{21} = - q_{22}=0.5.$ }\label{CIRfigures1}
\end{figure}

In this section we provide numerical examples to further interpret the optimal trading strategies. The regime-switching CIR model is capable of generating futures curves of different term structures. Figure \ref{Futures_Price} displays the futures prices in two different regimes. The CIR model with regime swicthing generates a convex curve in regime 1 (blue), and a concave curve in regime 2 (red).  In Figure \ref{CIRfigures1}, we illustrate the cases of optimal boundaries for futures trading under the CIR model in a two-regime market. As Figure \ref{CIRVVJJ0} shows, optimal boundaries divide the space into three disjoint regions in each regime, which can be specified as the long region (region below ``$\JJ$''), short region (region above ``$\VV$'') and the waiting region (region between ``$\JJ$'' and ``$\VV$''). The subscripts of value functions index to the regimes. Assuming the investor pre-commits to ``long-short'' strategy, it is optimal to take a long position first if the VIX is in the long region, and then exit the market when the VIX goes up to hit optimal boundary ``$\JJ$''. While if the investor adopts the ``short-long'' strategy, she will first short a futures and subsequently close out with a long position by the ``$\KK$'' and ``$\UU$'' boundaries in Figure \ref{CIRKKUU0} . In either case, our strategies confirm the intuition: ``buy low and sell high''. 

Figure \ref{CIRfigures1} also displays the optimal boundaries with different transaction costs. Without transaction costs (see left panel of Figure \ref{CIRfigures1}), the waiting region shrinks since the investor tends to enter and exit the market eariler, resulting in more rapid trades. In the presence of transaction costs (see right panel of Figure \ref{CIRfigures1}), the waiting region is widen in order to save on transaction costs. It should also be noted that as the transaction cost increases, the long boundary decreases and the short boundary increases, making the investor trade less frequently. In particular,  the fast divergence near expiry indicates that the investor should not enter the market after a critical time. The intuition is that a rational investor will never initiate a position if she does not have enough time to recover at least the transaction costs.

The ``$\PP$'' boundaries in Figure \ref{CIRfigures1} indicates the optimal value of VIX at which the investor should open a position. The boundary labeled as ``$\mathcal{P} = \AA$'' (resp. ``$\mathcal{P} = \BB$'') indicates the critical value at which the investor enters the market by taking a \textit{long} (resp. \textit{short}) futures position.  The investor should choose the ``short-first'' strategy if the VIX lies in the area above the ``$\mathcal{P} = \BB$'' boundary, whereas choose the ``long-first'' strategy if the VIX is lower than the ``$\mathcal{P} = \AA$'' boundary. The area  between the two boundaries is the region where the investor should wait for an better enter opportunity. This confirms our intuition -- take a long position when the VIX is low while take a short position when the VIX is high. Similarly, the waiting  region expands significantly near expiry to cover transaction costs. In other words, the investor will not enter the market unless the VIX is either very low or very high.

Once the investor finished choosing entry strategy, she could resort the exit strategy to corresponding optimal boundaries. For example, if the investor starts by longing a futures, then the optimal exit timing to close her position is represented by the optimal boundary ``$\VV$'' in Figure \ref{CIRfigures1}. However, if the investor's initial position is short, then she will hold the short position until the VIX hits ``$\UU$'' boundary.

\clearpage

\begin{figure}[h]
   \centering
\begin{subfigure}[b]{0.45\textwidth}
    \includegraphics[width=\textwidth]{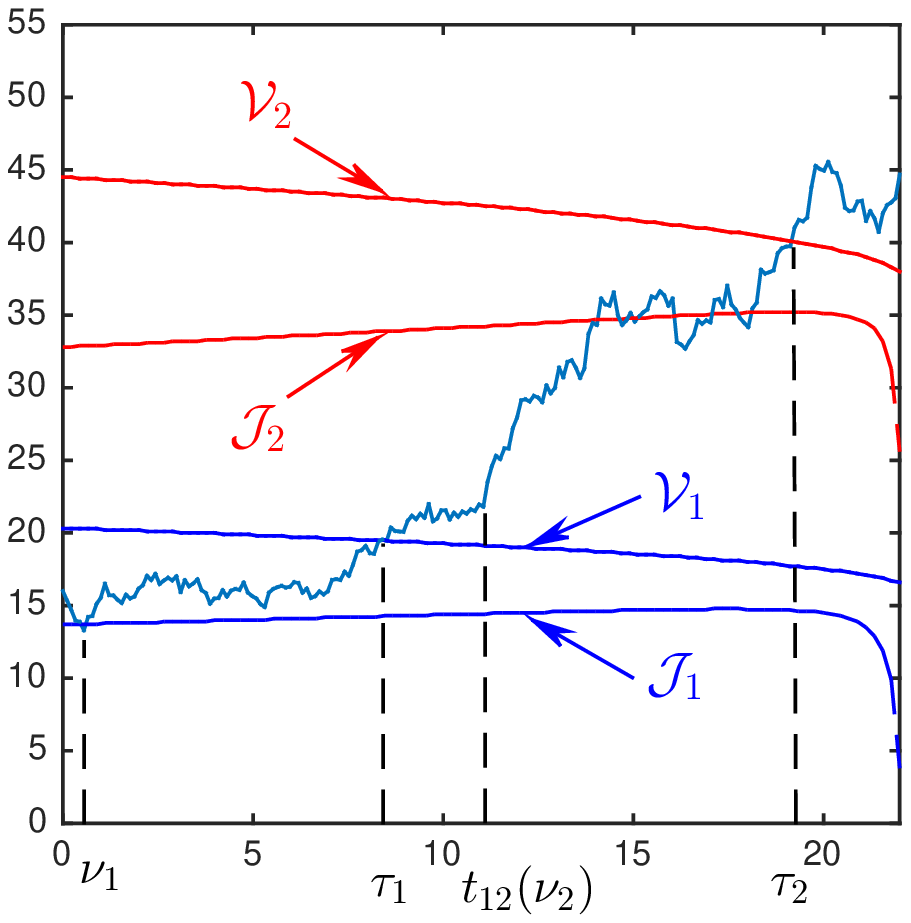}
     \caption{}\label{case1path1}
\end{subfigure}
\begin{subfigure}[b]{0.45\textwidth}
    \includegraphics[width=\textwidth]{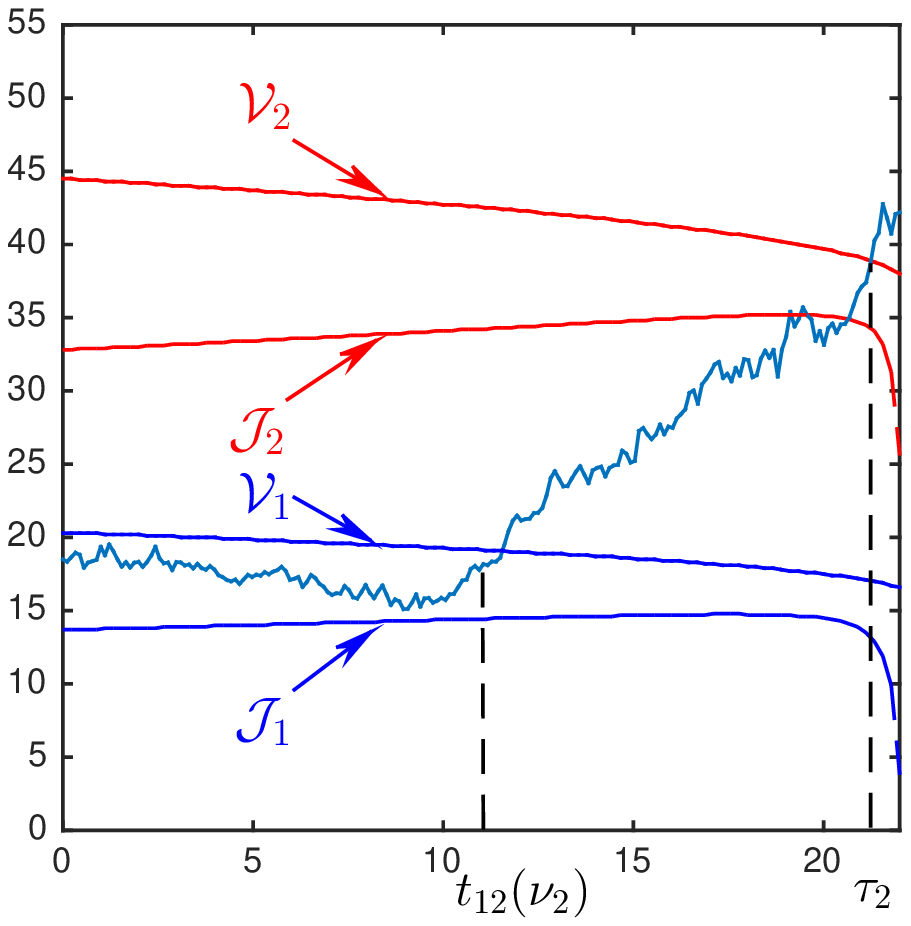}
    \caption{}\label{case1path2}
\end{subfigure}
\begin{subfigure}[b]{0.45\textwidth}
    \includegraphics[width=\textwidth]{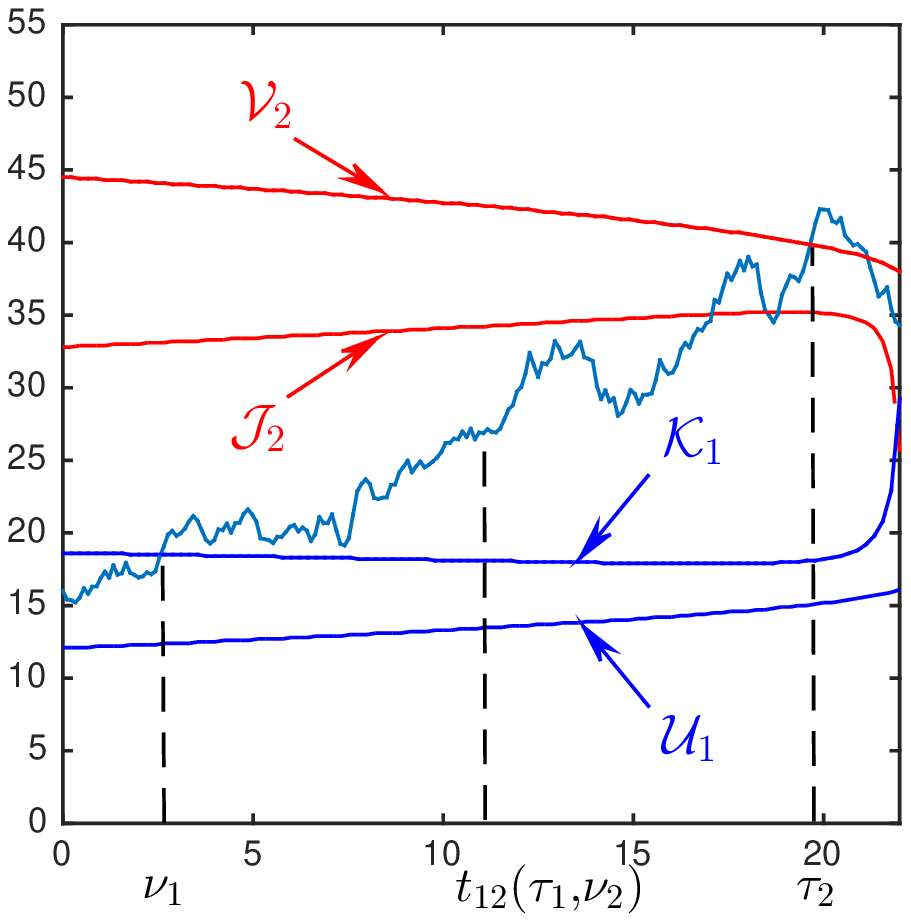}
    \caption{}\label{case1path3}
\end{subfigure}
 \caption{Simulated CIR paths and exercise times under 2-state regime-switching model.  (a) The investor enters at $\nu _1$ and exits at $\tau _1$ in regime 1; enters again at $\nu _2$ and exits at $\tau _2$ in regime 2.  (b) The investor enters at $\nu _2$ and exits at $\tau _2$ in regime 2 without any adjustments in regime 1. (c) The investor takes a short position at $\nu _1$ in regime 1 but switches to a long position at $t_{12}(\tau _1, \nu_2)$, and then liquilate her position at $\tau _2$ in regime 2. The parameters are the same as those in Figure \ref{CIRfigures1}. }\label{case1path}
\end{figure}

\clearpage

\begin{figure}
\centering
\begin{minipage}[t]{.5\textwidth}
\centering
\vspace{0pt}
\includegraphics[width=\textwidth]{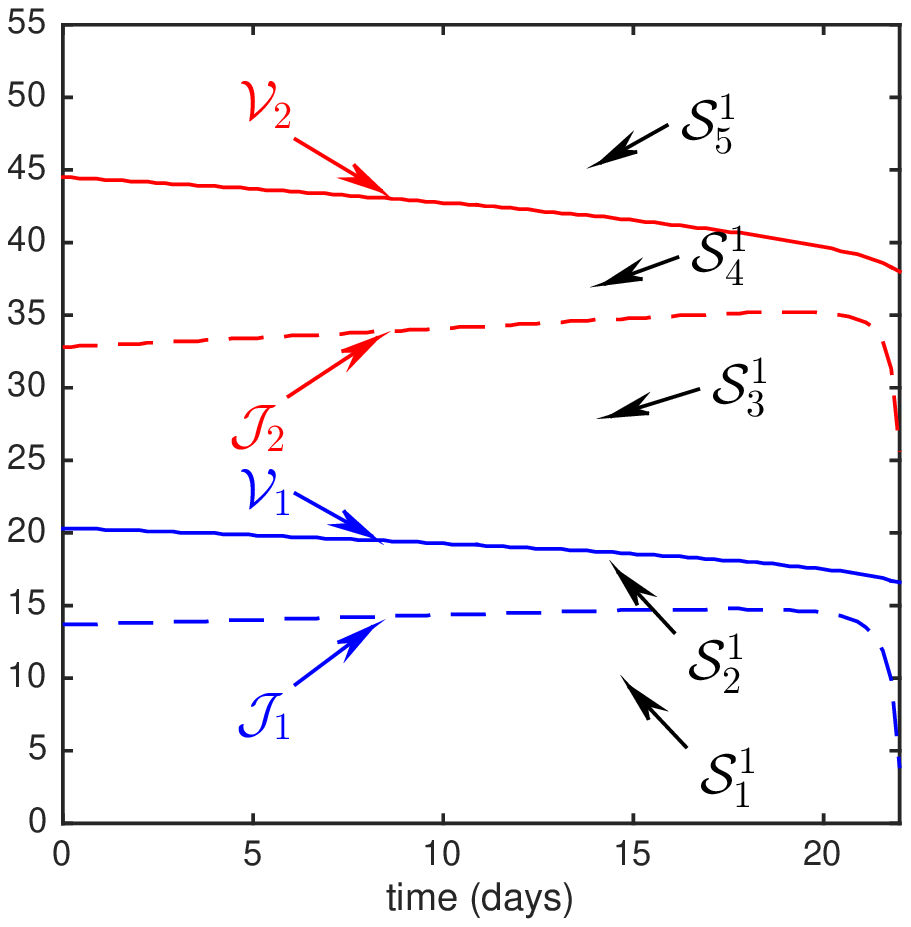}
\end{minipage}\hfill
\begin{minipage}[t]{.5\textwidth}
\centering
\vspace{0pt}
\captionof{table}{}
   \begin{tabular}{| c | c | c |}\hline
     &  In regime 1 & In regime 2 \\ \hline
      $S^1 _1$ &  Long & Long \\\hline
      $S^1 _2$ &  Wait & Long \\\hline
      $S^1 _3$ &  Short & Long \\\hline
      $S^1 _4$ &  Short & Wait \\\hline
      $S^1 _5$ & Short & Short \\ \hline
      \end{tabular}\label{Tcase1}
\end{minipage}
 \caption{Optimal boundaries for futures trading under the CIR model with 2-state regime switching. Parameters: $\hat{T}=\frac{22}{252} , T=\frac{66}{252}, r=0.05, \sigma_1=5.33, \sigma_2=6.42,  \theta_1= 17.58,  \theta_2= 39.5, \tilde{\theta}_1=18.16,  \tilde{\theta}_2=40.36, \mu _1=8.57, \mu _2=9, \tilde{\mu}_1=4.55, \tilde{\mu}_2=4.59, c=\hat{c}=0.01, q_{12}=-q_{11}= 0.1, q_{21} = - q_{22}=0.5.$}\label{case1}
\end{figure}

\begin{figure}
\centering
\begin{minipage}[t]{.5\textwidth}
\centering
\vspace{0pt}
\includegraphics[width=\textwidth]{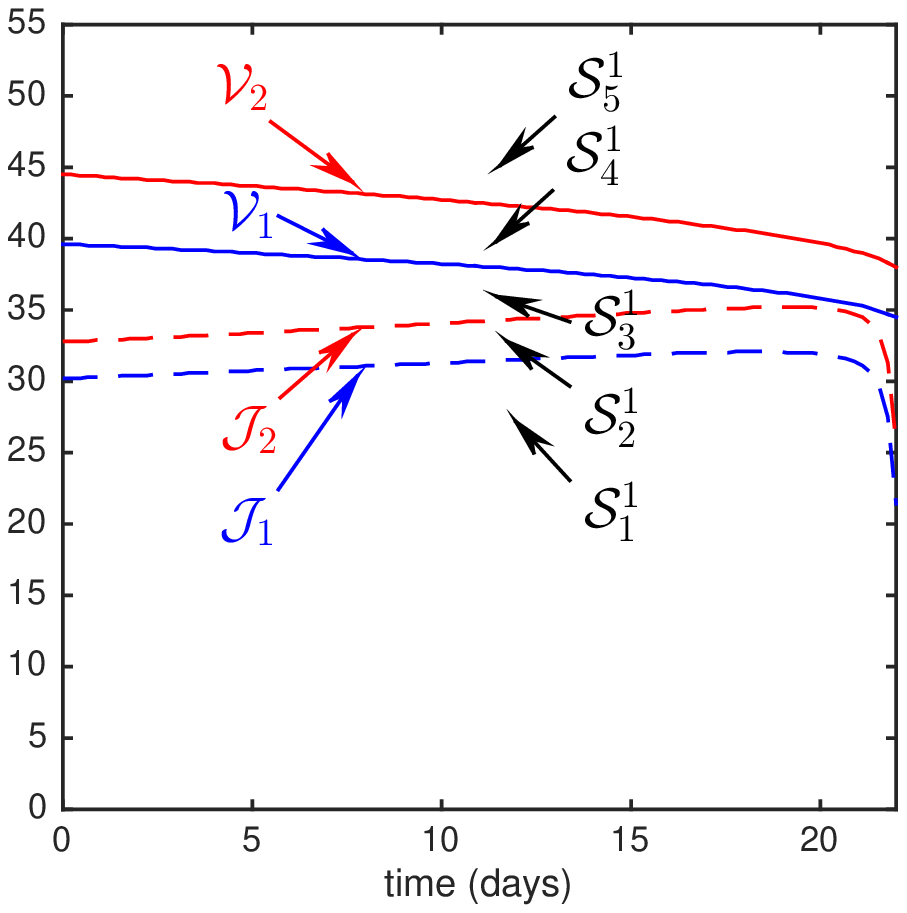}
\end{minipage}\hfill
\begin{minipage}[t]{.5\textwidth}
\centering
\vspace{0pt}
\captionof{table}{}
   \begin{tabular}{| c | c | c |}\hline
     &  In regime 1 & In regime 2 \\ \hline
      $S^1 _1$ &  Long & Long \\\hline
      $S^1 _2$ &  Wait & Long \\\hline
      $S^1 _3$ &  Wait & Wait \\\hline
      $S^1 _4$ &  Short & Wait \\\hline
      $S^1 _5$ & Short & Short \\ \hline
      \end{tabular}\label{Tcase2}
\end{minipage}
 \caption{Optimal boundaries for futures trading under the CIR model with 2-state regime switching. Parameters: $\hat{T}=\frac{22}{252} , T=\frac{66}{252}, r=0.05, \sigma_1=5.33, \sigma_2=6.42,  \theta_1= 35.6, \theta_2= 39.5, \tilde{\theta}_1=35.96,  \tilde{\theta}_2=40.36, \mu _1=8.57, \mu _2=9, \tilde{\mu}_1=4.55, \tilde{\mu}_2=4.59, c=\hat{c}=0.01, q_{12}=-q_{11}= 0.1, q_{21} = - q_{22}=0.5.$}\label{case2}
\end{figure}
\clearpage

To better explain our strategies, here we assume a two-state process and interpret the two regimes as a \textit{low-mean} regime (regime $1$) and a \textit{high-mean} regime (regime $2$). The regime-switching timing $t_{12}$ indicates the shift from the \textit{low-mean} regime to the \textit{high-mean} regime.  As shown in Figure \ref{case1path1}, the investor finishes the long-low-short-high tradings in both regimes. In the \textit{low-mean} regime, the investor chooses to long a futures position at time $\nu_1$ and then closes the position at $\tau_1$. When the regime switches at $t_{12}$, the VIX locates in the long region in \textit{high-mean} regime. The investor should long one position immediately at $t_{12}$ ($\nu _2$), and liquidate the position later at $\tau _2$ according to the optimal boundaries in regime 2.  Figure \ref{case1path2} shows another scenario. It is optimal for the investor to wait before the switch happens, since the VIX stays in the waiting region in regime $1$.  Then the investor would better to long a futures at $\nu_2$ and short at $\tau_2$ as the previous case.  Another example is shown in Figure \ref{case1path3}. The VIX goes up to hit the short boundary ``$\KK_1$'' at $\nu_1$ in regime 1.  The investor chooses  to short a position and she speculates that the price will decrease. However, the regime switches at time $t_{12}$ ($\tau_1$, $\nu_2$). The investor should immediately close her position and start to long one position according to the optimal boundaries ``$\JJ_2$'' and ``$\VV_2$''. The investor might need to face the loss in the position switching process. In other words, there is a positive probability of losses in a finite time period. We see that the regime-switching timing $t_{12}$ plays a key role for investor's trading decision. 

The introduction of regime switching adds considerable complexity to the optimal trading strategies. The last three examples in Figure \ref{case1path} are the simplest cases since we assume that the regime switching only happens once at $t_{12}$. Figure \ref{case1} and \ref{case2} show another approach to understand our models without specifying the regime-switching times. The process used here still has two states. Optimal boundaries separate the space into 5 regions. Table \ref{Tcase1} and \ref{Tcase2} indicate the exercises in different regimes by given VIX. Figure \ref{case1} shows the case of ``\textit{low-mean} regime vs \textit{high-mean} regime''. This regime-switching case captures some sudden changes of market (such as financial crisis in 2008), which might cause extremely high volatility. Once the regime switches, the investor is expected to take the long position aggressively. We can see the long region is as large as ``$S_1 ^1 + S_2 ^1 + S_3 ^1$'' in \textit{high-mean} regime. It shows that the investor has to adjust her strategies based on the different regimes, even though the VIX still stays in the same region. Figure \ref{case2} shows two regimes with relatively closer switching means. The optimal boundaries are higher in regime 2 than in regime 1, which means that the investor intends to enter and exit the market earlier in regime 1. Moreover, in regime 1, the waiting region is ``$S_2^1 + S_3^1$'', while in regime 2,  the waiting region is ``$S_3^1 + S_4^1$''. ``$S_3^1$'' is the common waiting region in both regimes. It implies that the investor might participate in tradings more frequently by adopting 2-regime model than single regime model. In other words, the presence of regime switching will impact investor's trading strategies. 

\begin{figure}[h]
   \centering
\begin{subfigure}[b]{0.48\textwidth}
    \includegraphics[width=\textwidth]{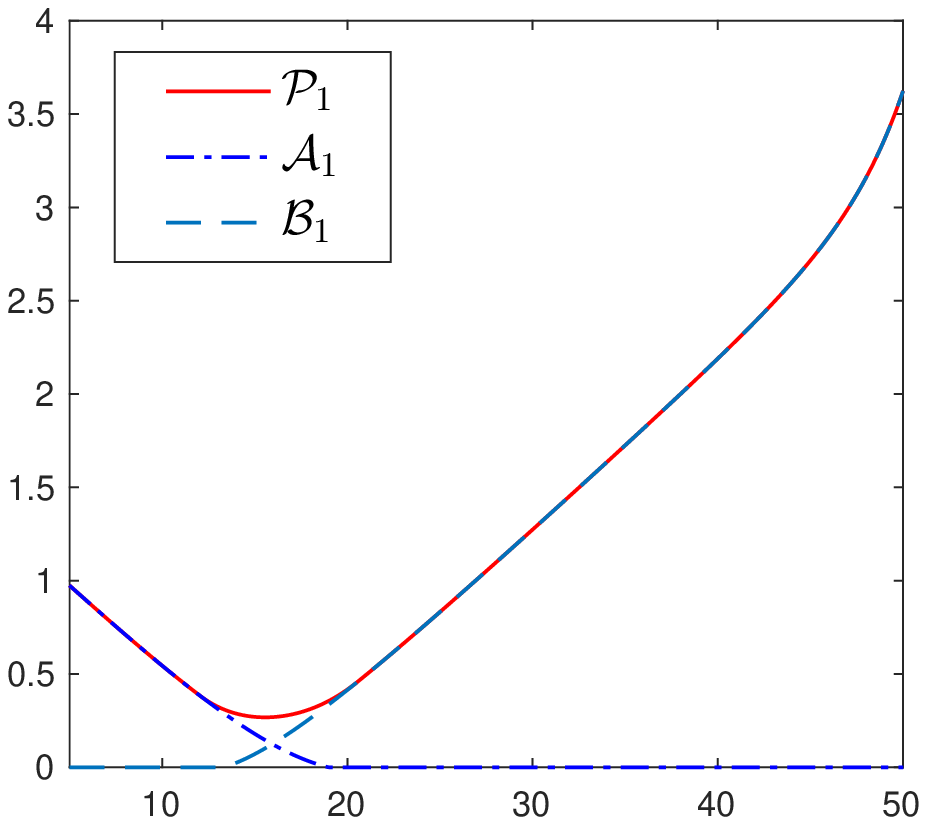}
     \caption{}\label{p1}
\end{subfigure}
\begin{subfigure}[b]{0.48\textwidth}
    \includegraphics[width=\textwidth]{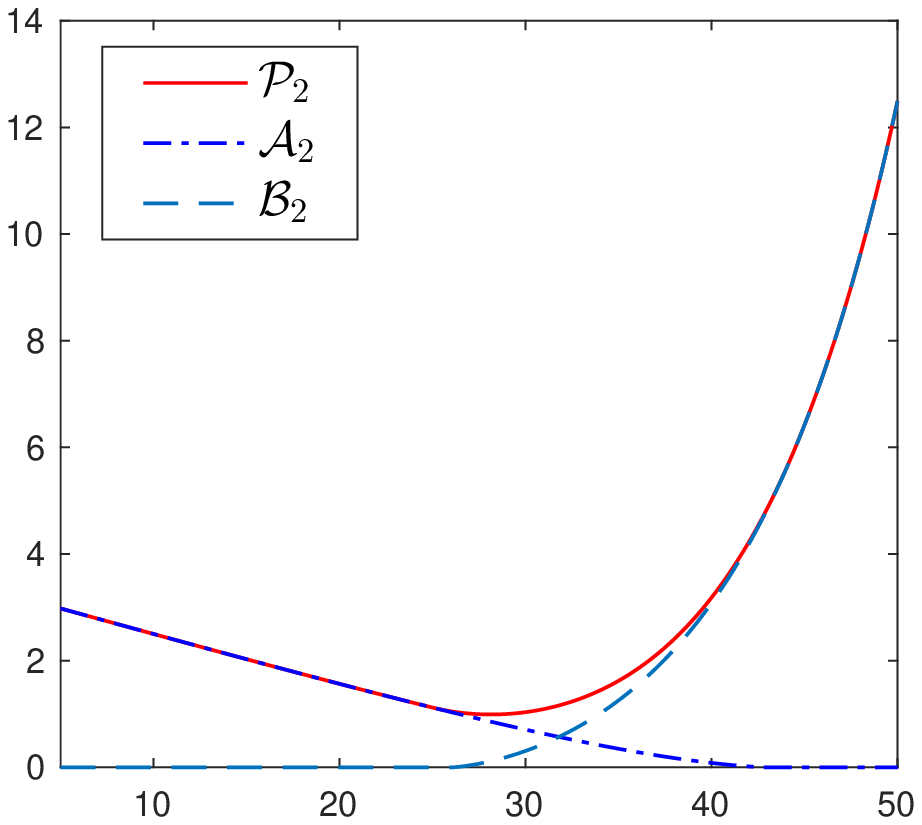}
    \caption{}\label{p2}
\end{subfigure}
 \caption{The value functions $\PP$, $\AA$, and $\BB$ in 2 regimes are plotted against the VIX at time 0. The parameters are the same as those in Figure \ref{CIRfigures1} .}\label{P}
\end{figure}

From the perspective of an investor with no position, she is interested in determining the best time to enter the market. We study the \emph{optimal timing premium} (see \cite{LeungLudkovski2011}, \cite{LeungLiu2012}, \cite{LeungLudkovski2}), which plays a vital role in the optimal strategies.  This premium  expresses the benefit of waiting to enter as compared to initialize the position  immediately. Precisely, the premium  is defined as 
\begin{align}\label{def_premium}
L(t,s,i) := \PP (t,s,i) - \max\{\AA(t, s, i), \BB(t,s,i)\}.
\end{align}
As we can see in \eqref{def_premium}, the optimal stopping time for $L(t,s, i)$ maximizes the expected discounted value from establishing the VIX futures position. Figure \ref{P} shows that $\PP$ dominates $\AA$ and $\BB$. We also note that $\PP = \AA$ when the VIX is low and $\PP = \BB$ when the VIX is high in each regime.

\section{Conclusion}
We extend the optimal VIX futures trading problems under a regime-switching model. This model allows the investor to capture the structural changes on the market. Numerical method is developed to solve these coupled system of variational inequalities that govern the value functions. Accounting for the  timing options as well as  the option to choose between a long or short position, we find that it is optimal to  delay market entry, as compared to the case of committing to either  go long or short \emph{a priori}. By introduce of regime-switching mechanism, it is noted that investor should modify her trading strategies correspondingly to regime-switching timings. The strategies and numerical method introduced in this paper can also be applied to other derivatives.

\bibliographystyle{apa}    
\bibliography{mybib2_05012016}     

\begin{thebibliography}{}

\bibitem[\protect\astroncite{Brennan and Schwartz}{1990}]{futuresBS}
Brennan, M.~J. and Schwartz, E.~S. (1990).
\newblock Arbitrage in stock index futures.
\newblock {\em Journal of Business}, 63(1):S7--S31.

\bibitem[\protect\astroncite{Buffington and Elliott}{2002}]{Buffington2002}
Buffington, J. and Elliott, R.~J. (2002).
\newblock American options with regime switching.
\newblock {\em International Journal of Theoretical and Applied Finance},
  5:497--514.

\bibitem[\protect\astroncite{Cartea et~al.}{2015}]{HFTbook}
Cartea, A., Jaimungal, S., and Penalva, J. (2015).
\newblock {\em Algorithmic and High-Frequency Trading}.
\newblock Cambridge University Press, Cambridge, England.

\bibitem[\protect\astroncite{Dai et~al.}{2011}]{futuresDaiKwok}
Dai, M., Zhong, Y., and Kwok, Y.~K. (2011).
\newblock Optimal arbitrage strategies on stock index futures under position
  limits.
\newblock {\em Journal of Futures Markets}, 31(4):394--406.

\bibitem[\protect\astroncite{Dotsis et~al.}{2007}]{dotsis2007empirical}
Dotsis, G., Psychoyios, D., and Skiadopoulos, G. (2007).
\newblock An empirical comparison of continuous-time models of implied
  volatility indices.
\newblock {\em Journal of Banking \& Finance}, 31(12):3584--3603.

\bibitem[\protect\astroncite{Elliott et~al.}{2008}]{elliott2008pde}
Elliott, R.~J., Siu, T.~K., and Chan, L. (2008).
\newblock A pde approach for risk measures for derivatives with regime
  switching.
\newblock {\em Annals of Finance}, 4(1):55--74.

\bibitem[\protect\astroncite{Gr\"{u}bichler and
  Longstaff}{1996}]{Grunbichler1996985}
Gr\"{u}bichler, A. and Longstaff, F. (1996).
\newblock Valuing futures and options on volatility.
\newblock {\em Journal of Banking and Finance}, 20(6):985--1001.

\bibitem[\protect\astroncite{Guo}{2001}]{guo2001explicit}
Guo, X. (2001).
\newblock An explicit solution to an optimal stopping problem with regime
  switching.
\newblock {\em Journal of Applied Probability}, pages 464--481.

\bibitem[\protect\astroncite{Khaliq and Liu}{2009}]{khaliq2009new}
Khaliq, A.~Q. and Liu, R. (2009).
\newblock New numerical scheme for pricing american option with
  regime-switching.
\newblock {\em International Journal of Theoretical and Applied Finance},
  12(03):319--340.

\bibitem[\protect\astroncite{Le and Wang}{2010}]{le2010finite}
Le, H. and Wang, C. (2010).
\newblock A finite time horizon optimal stopping problem with regime switching.
\newblock {\em SIAM Journal on Control and Optimization}, 48(8):5193--5213.

\bibitem[\protect\astroncite{Leung}{2010}]{Leung_regime2010}
Leung, T. (2010).
\newblock A {M}arkov-modulated stochastic control problem with optimal multiple
  stopping with application to finance.
\newblock In {\em 49th IEEE Conference on Decision and Control (CDC)}, pages
  559--566.

\bibitem[\protect\astroncite{Leung et~al.}{2016}]{LeungLiLiZheng2015}
Leung, T., Li, J., Li, X., and Wang, Z. (2016).
\newblock Speculative futures trading under mean reversion.
\newblock {\em Asia-Pacific Financial Markets}, pages 1--24.
\newblock Published online.

\bibitem[\protect\astroncite{Leung and Li}{2015}]{LeungLi2014OU}
Leung, T. and Li, X. (2015).
\newblock Optimal mean reversion trading with transaction costs and stop-loss
  exit.
\newblock {\em International Journal of Theoretical \& Applied Finance},
  18(3):15500.

\bibitem[\protect\astroncite{Leung and Li}{2016}]{meanreversionbook2016}
Leung, T. and Li, X. (2016).
\newblock {\em Optimal Mean Reversion Trading: Mathematical Analysis and
  Practical Applications}.
\newblock Modern Trends in Financial Engineering. World Scientific, Singapore.

\bibitem[\protect\astroncite{Leung et~al.}{2014}]{LeungLiWang2014CIR}
Leung, T., Li, X., and Wang, Z. (2014).
\newblock Optimal starting--stopping and switching of a {CIR} process with
  fixed costs.
\newblock {\em Risk and Decision Analysis}, 5(2):149--161.

\bibitem[\protect\astroncite{Leung et~al.}{2015}]{LeungLiWang2014XOU}
Leung, T., Li, X., and Wang, Z. (2015).
\newblock Optimal multiple trading times under the exponential {OU} model with
  transaction costs.
\newblock {\em Stochastic Models}, 31(4):554--587.

\bibitem[\protect\astroncite{Leung and Liu}{2012}]{LeungLiu2012}
Leung, T. and Liu, P. (2012).
\newblock Risk premia and optimal liquidation of credit derivatives.
\newblock {\em International Journal of Theoretical \& Applied Finance},
  15(8):1250059.

\bibitem[\protect\astroncite{Leung and Ludkovski}{2011}]{LeungLudkovski2011}
Leung, T. and Ludkovski, M. (2011).
\newblock Optimal timing to purchase options.
\newblock {\em SIAM Journal on Financial Mathematics}, 2(1):768--793.

\bibitem[\protect\astroncite{Leung and Ludkovski}{2012}]{LeungLudkovski2}
Leung, T. and Ludkovski, M. (2012).
\newblock Accounting for risk aversion in derivatives purchase timing.
\newblock {\em Mathematics \& Financial Economics}, 6(4):363--386.

\bibitem[\protect\astroncite{Leung and Shirai}{2015}]{LeungShirai}
Leung, T. and Shirai, Y. (2015).
\newblock Optimal derivative liquidation timing under path-dependent risk
  penalties.
\newblock {\em Journal of Financial Engineering}, 2(1):1550004.

\bibitem[\protect\astroncite{Leung and Yamazaki}{2013}]{LeungYamazakiQF}
Leung, T. and Yamazaki, K. (2013).
\newblock {A}merican step-up and step-down credit default swaps under
  {L}{\'e}vy models.
\newblock {\em Quantitative Finance}, 13(1):137--157.

\bibitem[\protect\astroncite{Menc{\'i}a and Sentana}{2013}]{futuresVIX}
Menc{\'i}a, J. and Sentana, E. (2013).
\newblock Valuation of {VIX} derivatives.
\newblock {\em Journal of Financial Economics}, 108(2):367--391.

\bibitem[\protect\astroncite{Sircar and
  Papanicolaou}{2014}]{futures_SircarAndrew}
Sircar, R. and Papanicolaou, A. (2014).
\newblock A regime-switching {H}eston model for {VIX} and {S}\&{P} 500 implied
  volatilities.
\newblock {\em Quantitative Finance}, 14(10):1811--1827.

\bibitem[\protect\astroncite{Stepanek}{2015}]{stepanek2015comparison}
Stepanek, C. (2015).
\newblock Comparison of commodity future pricing approaches with cointegration
  techniques.
\newblock {\em Journal of Financial Engineering}, 2(01):1550002.

\bibitem[\protect\astroncite{Wang and Daigler}{2011}]{wangVIX2011}
Wang, Z. and Daigler, R.~T. (2011).
\newblock The performance of {VIX} option pricing models: Empirical evidence
  beyond simulation.
\newblock {\em Journal of Futures Markets}, 31(3):251--281.

\bibitem[\protect\astroncite{Whaley}{2000}]{whaley2000investor}
Whaley, R.~E. (2000).
\newblock The investor fear gauge.
\newblock {\em The Journal of Portfolio Management}, 26(3):12--17.

\bibitem[\protect\astroncite{Wilmott et~al.}{1995}]{wilmottbook1995}
Wilmott, P., Howison, S., and Dewynne, J. (1995).
\newblock {\em The Mathematics of Financial Derivatives: A Student
  Introduction}.
\newblock Cambridge University Press, 1st edition.

\bibitem[\protect\astroncite{Yao et~al.}{2006}]{yao2006regime}
Yao, D.~D., Zhang, Q., and Zhou, X.~Y. (2006).
\newblock A regime-switching model for {E}uropean options.
\newblock In {\em Stochastic processes, optimization, and control theory:
  applications in financial engineering, queueing networks, and manufacturing
  systems}, pages 281--300. Springer.

\bibitem[\protect\astroncite{Zhang and Zhu}{2006}]{futures_zhang}
Zhang, J.~E. and Zhu, Y. (2006).
\newblock {VIX} futures.
\newblock {\em Journal of Futures Markets}, 26(6):521--531.

\end{thebibliography}
 
\end{document}